\documentclass[twoside,12pt]{article}
\usepackage{epsfig}

\def\Journal#1#2#3#4{{#1} {#2} (#4) #3 }

\def\NPA{{\em Nucl. Phys.} A}

\def\PRD{{\em Phys. Rev.} D}

\newcommand{\be}{\begin{equation}}
\newcommand{\ee}{\end{equation}}
\newcommand{\bea}{\begin{eqnarray}}
\newcommand{\eea}{\end{eqnarray}}

\begin{document}

\title{ \vspace{1cm} Aspects of the phase diagram in (P)NJL-like
models}
\author{M.\ Buballa,$^1$
A.G.\ Grunfeld,$^{2,3}$ 
A.E.\ Radzhabov,$^4$
D. \ Scheffler$^1$\\ 
\\
$^1$Institut f\"ur Kernphysik, Technische Universit\"at Darmstadt, Germany\\
$^2$CONICET, Buenos Aires, Argentina\\
$^3$Physics Dept., Comisi\'on Nacional de Energ\'{\i}a At\'omica,  
Buenos Aires, Argentina\\
$^4$Institute for System Dynamics and Control Theory, Irkutsk, Russia
}
\maketitle
\begin{abstract} 
We discuss three applications of NJL- and PNJL-like models 
to assess aspects of the QCD phase diagram:
First, we study the effect of mesonic correlations on the pressure
below and above the finite temperature phase transition within
a nonlocal PNJL model beyond the mean-field approximation. 
Second, we reconstruct the phase boundary of an NJL model from a Taylor 
expansion of the chiral susceptibility about $\mu = 0$
and compare the result with the exact phase boundary.
Finally, we demonstrate the realization of the ``non-standard scenario''
for the critical surface in a three-flavor PNJL model with a 
$\mu$-dependent determinant interaction.

\end{abstract}
\section{Introduction}

The investigation of the QCD phase transition and the 
structure of the phase diagram is certainly one of the most exciting 
areas in the field of strong interaction physics. 
A theoretical approach to these questions requires
non-perturbative methods, which also provide a proper understanding
of the chiral quark dynamics and the confinement mechanism.
Until now, the only method which is directly based on QCD and which
meets these requirements is lattice gauge theory.
Unfortunately, the application of lattice results to experimental data is
complicated by the fact that most lattice calculations are often
performed with rather large quark masses, leading to unphysically
large pion masses. Even more difficult is the extension of lattice
calculations to nonvanishing chemical potentials, where
the standard importance sampling techniques are spoiled by the 
``sign problem''. 
To bridge this gap between existing lattice data and 
phenomenologically interesting regimes which are not yet accessible
by first principles, effective models which share the relevant
symmetries with QCD can deal as useful tools.
In this context, the Nambu--Jona-Lasinio (NJL) model is probably the
most popular example. More recently, the Polyakov-loop extended 
Nambu--Jona-Lasinio (PNJL) \cite{Meisinger:1995ih,Fukushima:2003fw}
has received increasing interest,
as it allows to investigate both, the chiral and the deconfinement phase 
transition by studying the corresponding order parameters.
Moreover, the coupling of the quark dynamics to the Polyakov loop 
cures, at least to a large extent, one of the most disturbing features
of the NJL model, namely the contribution of unconfined quarks to the 
pressure at low temperatures and densities. 

In the following, we will discuss three applications of NJL and PNJL-models
to issues related to the phase diagram.

\section{Effects of mesonic correlations}

In spite of the simplicity of the PNJL model, a remarkable agreement
with the results of lattice QCD thermodynamics have been obtained 
in Ref.~\cite{Ratti:2005jh}.
However, that comparison was not entirely consistent:
Whereas unphysically large values of the current quark masses have been
used in the lattice simulations, physical values have been employed in 
the PNJL model analysis of Ref.~\cite{Ratti:2005jh}.
Moreover, after successfully removing (most of) the unphysical quark degrees
of freedom from the confined phase, the PNJL model treated in mean-field
approximation as in Ref.~\cite{Ratti:2005jh} does not contain {\it any}
degree of freedom in this regime.
Obviously, this is a rather poor description of the hadronic phase at finite
temperature where mesons are expected to become relevant.
Hence, the good agreement of the PNJL results with the lattice data could
be partially accidental in that mesonic correlations have been neglected in
the PNJL analysis, while in the lattice calculations they are suppressed by
the large current quark masses.
To get a consistent picture one should thus go beyond the mean-field
approximation and include mesonic correlations.

Here we briefly discuss the main results of Ref.~\cite{Blaschke:2007np},
where we have done this in the framework of a $1/N_c$ expansion to 
next-to-leading order. 
Up to an additive constant, the thermodynamic potential at $\mu=0$
is then given by 
$\Omega(T) = \Omega_\mathrm{mean\;field}(T) + \Omega_\mathrm{corr}(T)$
where the correlation part
\be
\Omega_{\mathrm{corr}}(T) 
= 
\sum\limits_{M=\pi,\sigma} 
            d_M\,\frac{T}{2}
            \sum\limits_m \int\frac{d^3p}{(2\pi)^3} 
            \ln\big[1-G\Pi_M(\vec p,\omega_m)\big]
\ee
corresponds to the ring-sum of RPA-like polarization loops $\Pi_M$
in the mesonic channel $M$ with degeneracy factor $d_M$.
These terms are worked out within a two-flavor PNJL-type model 
with a nonlocal 4-point interaction in the quark sector with a
Gaussian form factor.
This interaction has the advantage that all diagrams are finite.
Further details of the model can be found in Ref.~\cite{Blaschke:2007np}.

\begin{figure}[tb]
\begin{center}
\begin{minipage}[t]{16.5 cm}
\epsfig{file=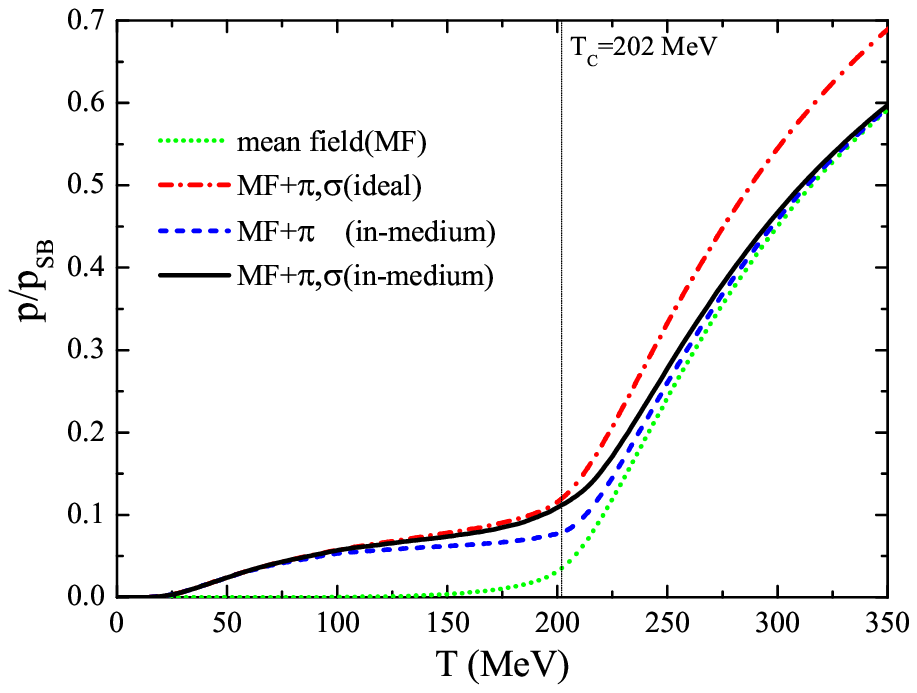,scale=0.8}
\quad
\epsfig{file=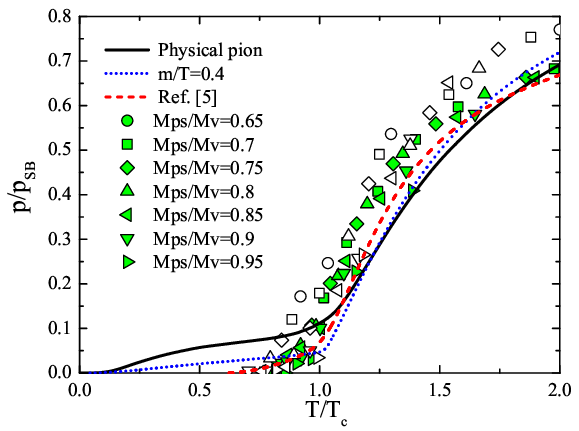,scale=1.29}
\end{minipage}
\begin{minipage}[t]{16.5 cm}
\caption{Left: Scaled pressure $p/p_{SB}$ 
in the nonlocal PNJL model with the physical pion mass:
mean field contribution (green dotted line),
mean field $+$ pion (blue dashed line),
mean field $+$ pion $+$ sigma (black solid line).
The red dash-dotted line denotes the scaled pressure of an ideal
pion $+$ sigma gas with fixed masses.
Right:
Scaled pressure $p/p_{SB}$ as a function of $T/T_c$:
non-local PNJL model with physical pion mass (black solid line)
and with $m_c/T=0.4$ (blue dotted line).
Red dashed line: Lattice data for two-flavor QCD with
staggered quarks ~\cite{Karsch:2000ps}. 
Points: Lattice data for two-flavor QCD with
Wilson-type quarks ~\cite{AliKhan:2001ek} for $N_t=6$ (open symbols) and
$N_t=4$ (green filled symbols). 
The data for the pressure \cite{AliKhan:2001ek}
have been divided by the Stefan-Boltzmann limit for $N_t=6$ and $N_t=4$, 
respectively, as given in Ref.~\cite{AliKhan:2001ek}.
Figures adapted from Ref.~\cite{Blaschke:2007np}.
\label{fig:pcorr}
}
\end{minipage}
\end{center}
\end{figure}

Our main results are shown in Fig.~\ref{fig:pcorr}.
On the left hand side, the pressure $p(T) = -\Omega(T)$,
divided by the Stefan-Boltzmann limit, is displayed as a function
of the temperature. 
For comparison with the full result
we also show the mean-field result and the
mean-field plus pion contribution as well as the
result for an ideal pion and sigma gas with the masses fixed at
their vacuum values.
We find that
at low temperatures the mean-field (i.e., quark) contribution is suppressed
and the pressure can be well described by a free pion gas.
Near the critical temperature the $\sigma$ meson gives an additional visible
contribution whereas already for $T>1.5~T_c$ the mesonic contributions are
negligible and the quark-gluon mean-field dominates the pressure.

The results shown in the left panel of Fig.~\ref{fig:pcorr} have been
obtained using realistic parameters with a current quark mass 
$m = 5.8$~MeV, corresponding to the physical pion 
mass of 140~MeV in vacuum. 
On the other hand, in most lattice calculations the masses are considerably 
larger. Thus in order to perform a meaningful comparison with lattice
results, we should repeat our model calculations using quark masses
similar to the lattice ones. 
To be specific, we choose the current quark mass to scale with the
temperature as $m = 0.4~T$, mimicking the situation in the lattice
calculation of Ref.~\cite{Karsch:2000ps}, where the up and down quark
masses behave in this way.
The resulting pressure is displayed by the blue dotted line in the right
panel of Fig.~\ref{fig:pcorr}. For comparison we show again the result 
obtained with $m = 5.8$~MeV  (black solid line). 
With $m = 0.4~T$ the qualitative behavior remains unchanged.
Quantitatively, the meson contributions are of course suppressed
by their higher masses, but they are still visible.

The red dashed line indicates the result of Ref.~\cite{Karsch:2000ps}
obtained on a $16^3\times 4$ lattice with improved staggered fermion actions
for two flavors. As mentioned above, in this calculation the quark mass 
scales with the temperature, $m_{u,d}/T = 0.4$ and should therefore be
compared with the blue dotted line.
In the figure, we also show the lattice results of
Ref.~\cite{AliKhan:2001ek} for Wilson fermions.
Although these data have been extracted
for $T$-independent quark masses and are therefore not entirely comparable 
to the staggered fermion data (and to our calculations), they show a 
similar tendency.

Whereas in our model, even for $m = 0.4~T$, there remains a visible pion 
contribution to the pressure down to $T/T_c \approx 0.2$, 
the lattice pressure of \cite{Karsch:2000ps} vanishes at $T/T_c = 0.6$.
This discrepancy has a trivial explanation by the fact that the lattice
pressure has been obtained by the ``integral method'' which leaves one
integration constant undetermined.
In Ref.~\cite{Karsch:2000ps}, this constant has been fixed by the choice
that the pressure vanishes at $T/T_c = 0.6$.
On the other hand, chiral perturbation theory predicts that at very low
temperature the pressure is well described by an ideal pion gas,
in good agreement with our model.
Hence, if the integration constant on the lattice had been fitted to
$\chi PT$, rather than setting the pressure at $T = 0.6~T_c$ equal to
zero, there would be good agreement with our results at this point.
On the other hand, our approach underestimates the lattice
data in the region between about 0.9 and 1.6~$T_c$.
The missing pressure would be even larger if we shift the 
lattice pressure upwards to obtain agreement at $T = 0.6~T_c$.
This indicates that hadronic resonances, other than pion and sigma
become important in this regime and should be taken into account in 
improved versions of the model.

\section{Assessing the phase diagram via Taylor expansion}

As pointed out in the introduction a straight forward application of
standard lattice techniques to the regime of non-vanishing 
quark number chemical potential $\mu$ is not possible. 
In this context various methods have been suggested to circumvent 
these problems. 
One of these methods consists in performing a Taylor expansion 
of the pressure in powers of $\mu/T$,
with the Taylor coefficients evaluated at $\mu = 0$,
\be
   \frac{p}{T^4}(T,\mu) = 
     \sum_{n=0}^{\infty} c_n(T) \left(\frac{\mu}{T}\right)^n\,,
   \qquad
    c_n(T) = \frac{1}{n!} \left.
     \frac{\partial}{\partial(\mu/T)^n}
     \left(\frac{p}{T^4}(T,\mu)\right)\right|_{\mu=0}\,.
\label{pTe}
\ee
One may also consider different chemical potentials for up, down, and 
strange quarks or, equivalently, for baryon number, electric charge and 
strangeness, but here we want to restrict ourselves to the case of a 
single chemical potential.
Because of charge conjugation symmetry, only even powers in $\mu/T$ appear.
In recent lattice calculations coefficients up to $n=8$ have been
calculated \cite{Schmidt:2008cf}, but most studies so far were restricted to 
$n \leq 6$ \cite{Allton:2005gk}.
It is then a natural question, how reliably this information can be used to
reconstruct the phase boundary at non-zero $\mu$. It would also be 
interesting to know whether one could find signatures of a critical endpoint 
in this way. Such questions can be addressed by performing a Taylor
expansion in a model where the phase boundary and the location of the 
critical endpoint are known ``exactly''
(in the sense of the model). 

\begin{figure}[tb]
\begin{center}
\begin{minipage}[t]{16.5 cm}
\epsfig{file=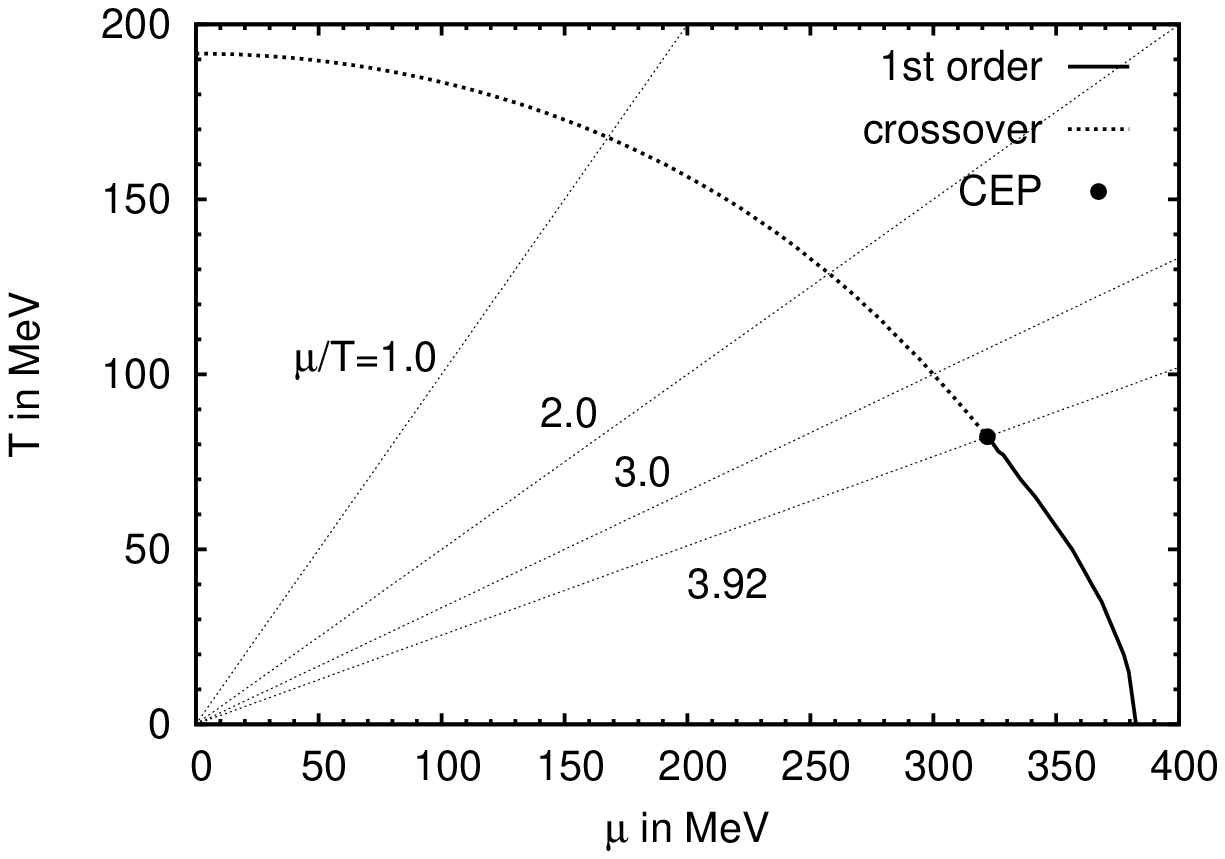,scale=0.65}
\epsfig{file=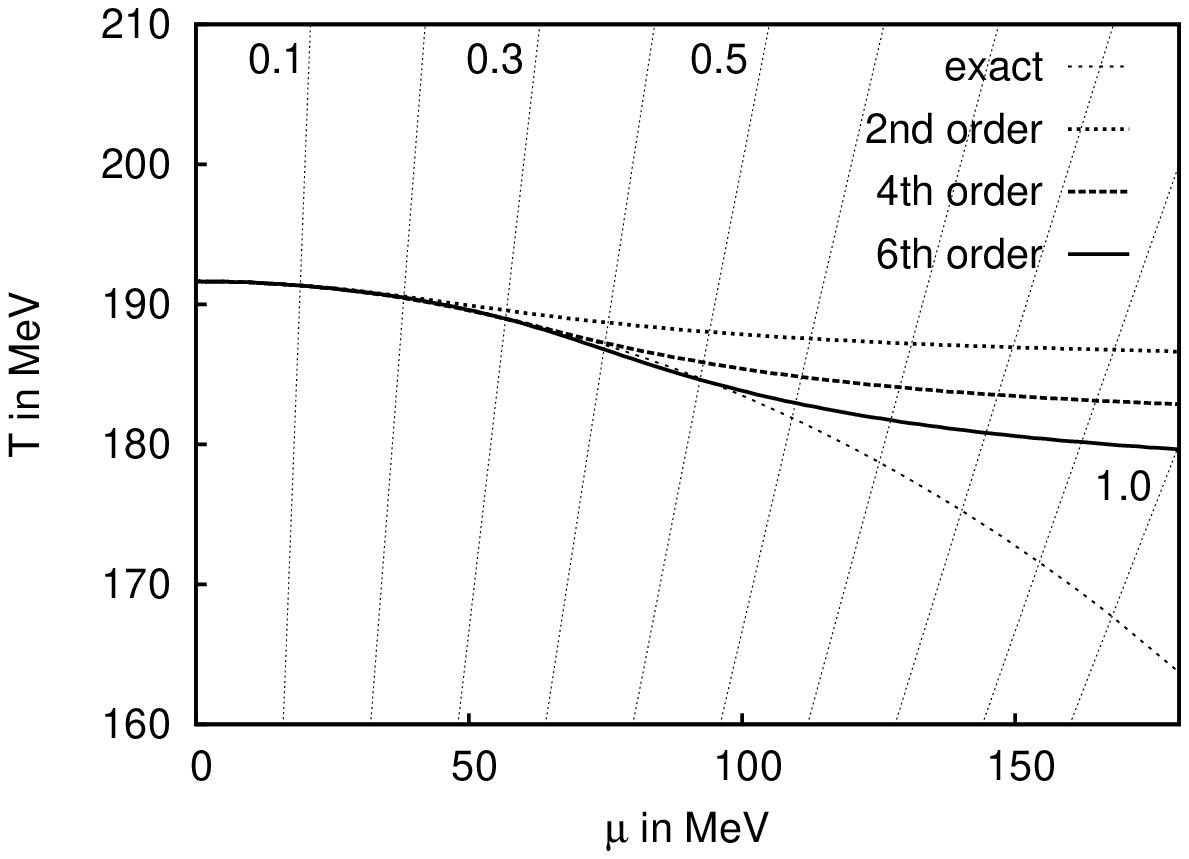,scale=0.65}
\end{minipage}
\begin{minipage}[t]{16.5 cm}
\caption{Phase diagram of the 2-flavor NJL model. Left: exact mean-field
result for the crossover line (dotted) and the first-order phase boundary
(solid). 
Right: comparison of the exact crossover line at small chemical
potential with the reconstructed results from the Taylor expansion of the
chiral susceptibility to 2nd, 4th, and 6th order. 
In both figures, the radial lines are lines of constant $\mu/T$, as
indicated. Adapted from Ref.~\cite{SchefflerBSc}.
\label{fig:taylorpd}}
\end{minipage}
\end{center}
\end{figure}

Here we want to discuss some results of Ref.~\cite{SchefflerBSc}, where
the Taylor expansion has been studied within a two-flavor NJL model
without Polyakov loop in mean-field approximation. 
The ``exact'' phase diagram of that model is shown in the left panel of 
Fig.~\ref{fig:taylorpd}. At large chemical potentials we find a 
first-order phase boundary (solid line), which ends at a critical point 
at $\mu/T \approx 4$. For lower values of $\mu/T$, there is a smooth
crossover, which is indicated by the dotted line. This crossover line
has been defined as the maxima of the reduced chiral susceptibility,
\be
   \frac{\chi_{mm}(T,\mu)}{T^2} 
   = \frac{1}{T^2} \frac{\partial^2 p(T,\mu)}{\partial m^2}\,,
\label{rcs}
\ee
along lines of constant $\mu/T$. Here $m$ is the current quark mass 
parameter of the NJL model.
Eq.~(\ref{rcs}) can be Taylor expanded as well. 
Employing Eq.~(\ref{pTe}) we obtain
\be
   \frac{\chi_{mm}(T,\mu)}{T^2}
= 
     \sum_{n=0}^{\infty} c^\chi_n(T) \left(\frac{\mu}{T}\right)^n\,,
   \qquad
   c_n^\chi = T^2 \frac{\partial^2 c_n}{\partial m^2}\,.
\ee

\begin{figure}[tb]
\begin{center}
\begin{minipage}[t]{16.5 cm}
\epsfig{file=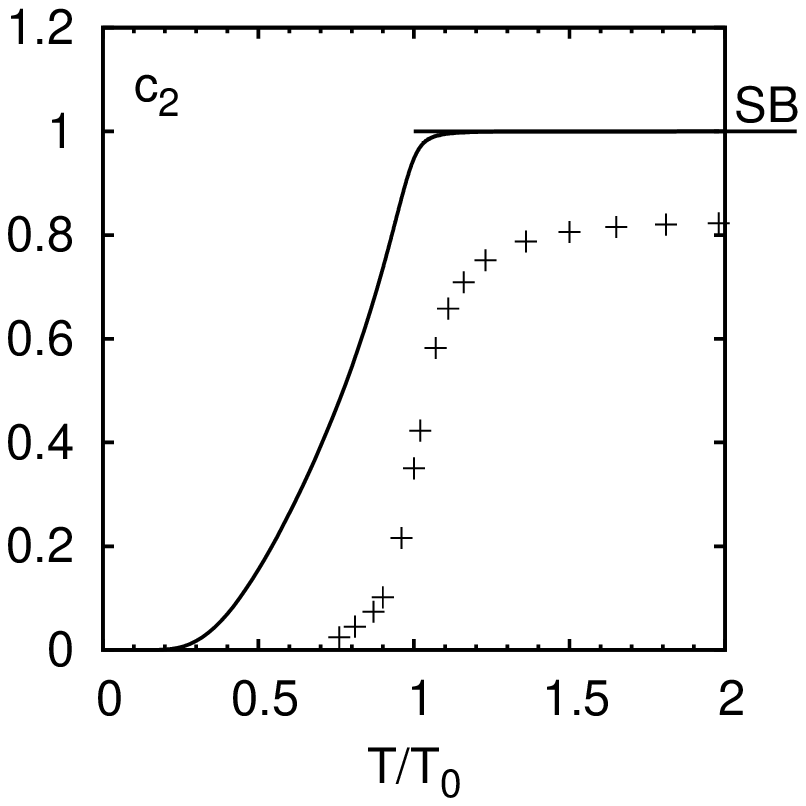,scale=0.5}
\hspace{-1.5cm}
\epsfig{file=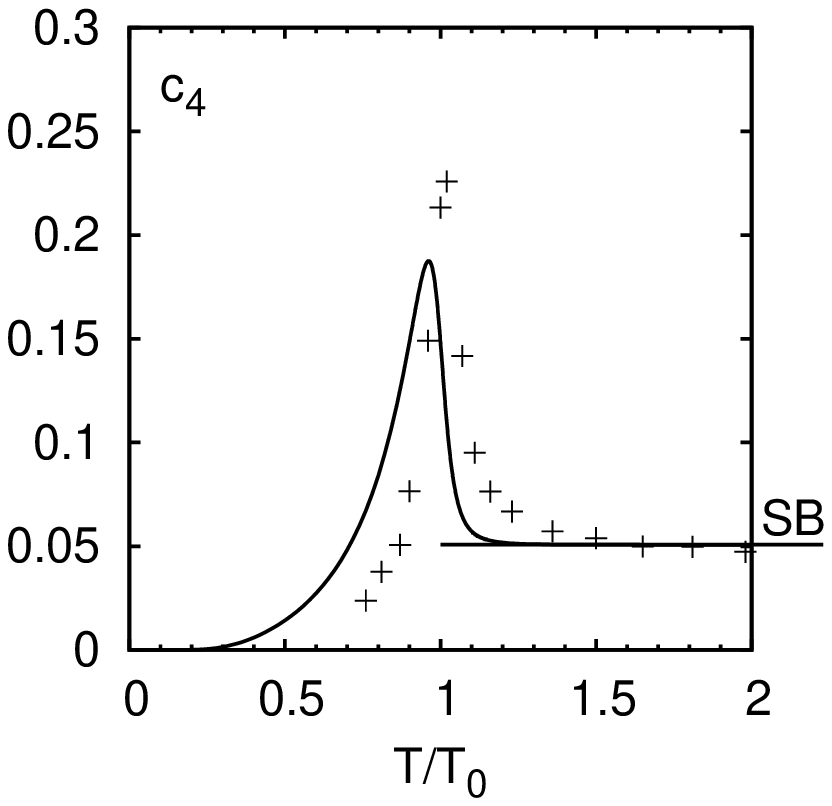,scale=0.5}
\hspace{-1.5cm}
\epsfig{file=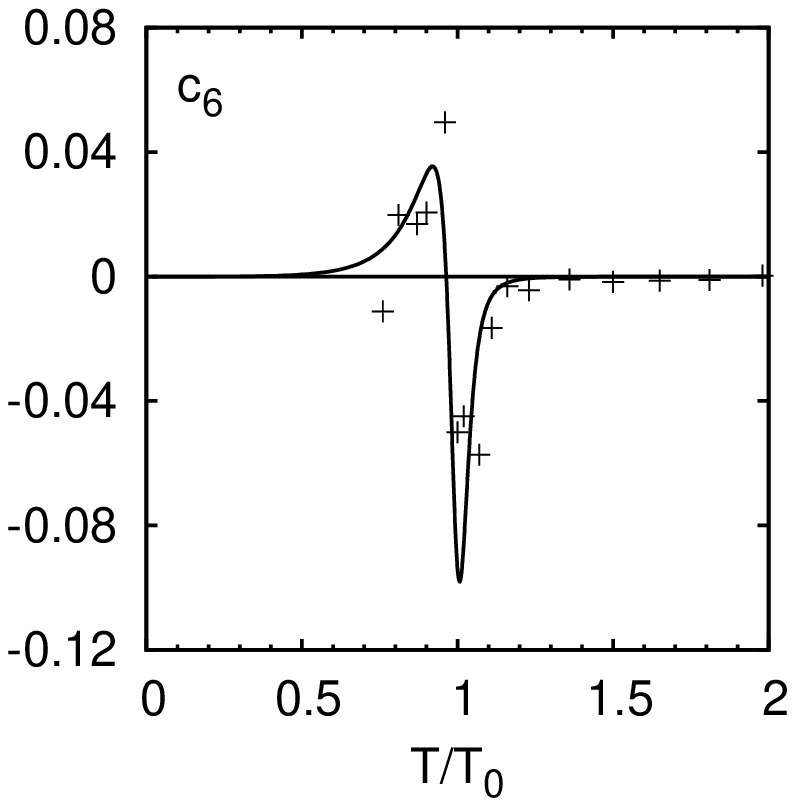,scale=0.5}
\epsfig{file=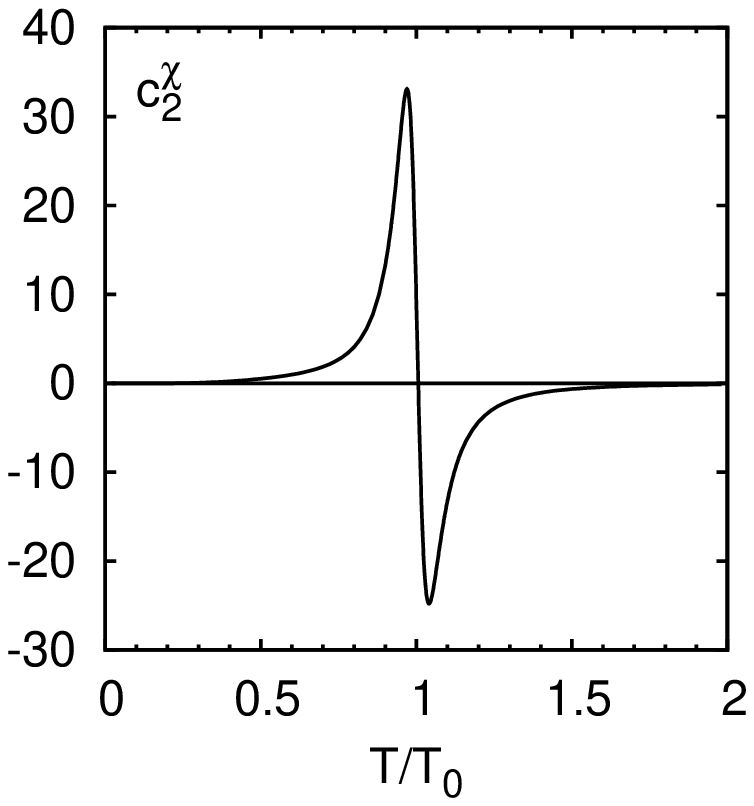,scale=0.5}
\hspace{-1.5cm}
\epsfig{file=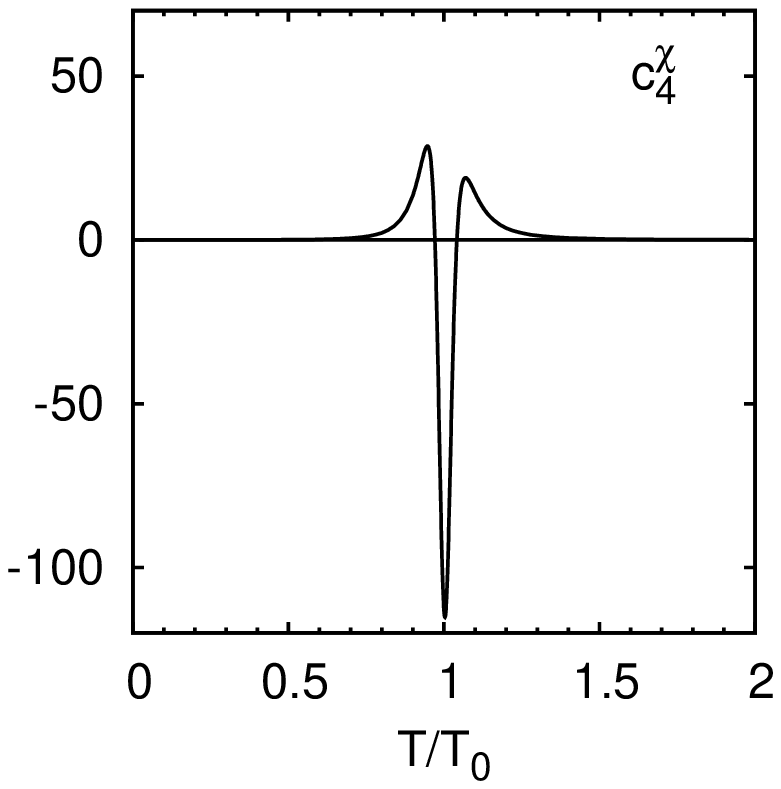,scale=0.5}
\hspace{-1.5cm}
\epsfig{file=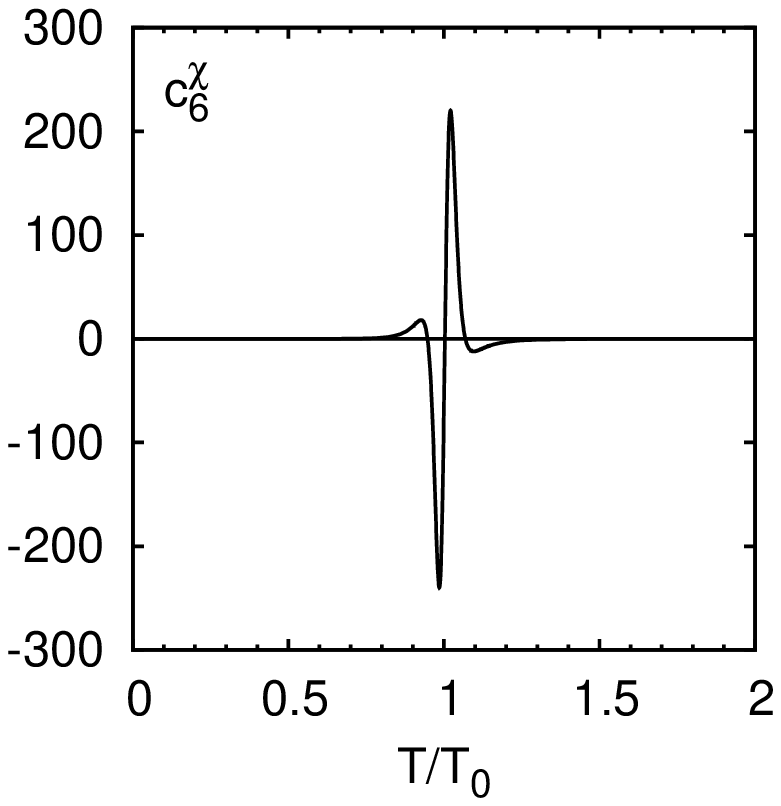,scale=0.5}
\end{minipage}
\begin{minipage}[t]{16.5 cm}
\caption{Taylor expansion coefficients $c_n$ (upper panels) and
$c^\chi_n$ (lower panels) for $n = 2,4,6$ as functions of $T/T_0$
(lines). Also shown are lattice results for $c_n$ from 
Ref.~\cite{Allton:2005gk} (points). Adapted from Ref.~\cite{SchefflerBSc}.
\label{fig:taylorcoeff}}
\end{minipage}
\end{center}
\end{figure}

In Fig.~\ref{fig:taylorcoeff} the coefficients $c_n$ and $c^\chi_n$ for 
$n = 2,4,6$ are displayed as functions of $T/T_0$, where $T_0$ is the 
crossover temperature at $\mu = 0$. Our model results are indicated by
the lines. For the $c_n$ we also show the lattice results of 
Ref.~\cite{Allton:2005gk} for illustration. We see that the qualitative
behavior is similar, but there are quantitative differences. In 
particular the Stefan Boltzmann limits, which are also indicated in the 
figure, are reached much faster in the model. 
This is an artifact of the NJL model which can be improved considerably
by including the Polyakov loop \cite{Rossner:2007ik}.

However, our present focus is not on a comparison between model and 
lattice calculations but on a comparison within the NJL model between the 
``exact'' phase boundary and the reconstructed result obtained from the
chiral susceptibility calculated with the first few Taylor terms.
This comparison is shown in the right panel of Fig.~\ref{fig:taylorpd}. 
As one can see, even with the 6th-order terms included, the crossover 
line starts to deviate from the exact result already below 
$\mu/T \approx 0.6$.
This limit cannot be pushed much further by including more Taylor
terms: As obvious from Fig.~\ref{fig:taylorcoeff}, higher-order Taylor 
terms become more and more oscillatory. As a consequence, the reconstructed
chiral susceptibility does not have one, but several maxima above a certain
value of $\mu/T$ and the crossover line is no longer unique. 

\section{The critical surface}

\begin{figure}[tb]
\begin{center}
\begin{minipage}[t]{16.5 cm}
\epsfig{file=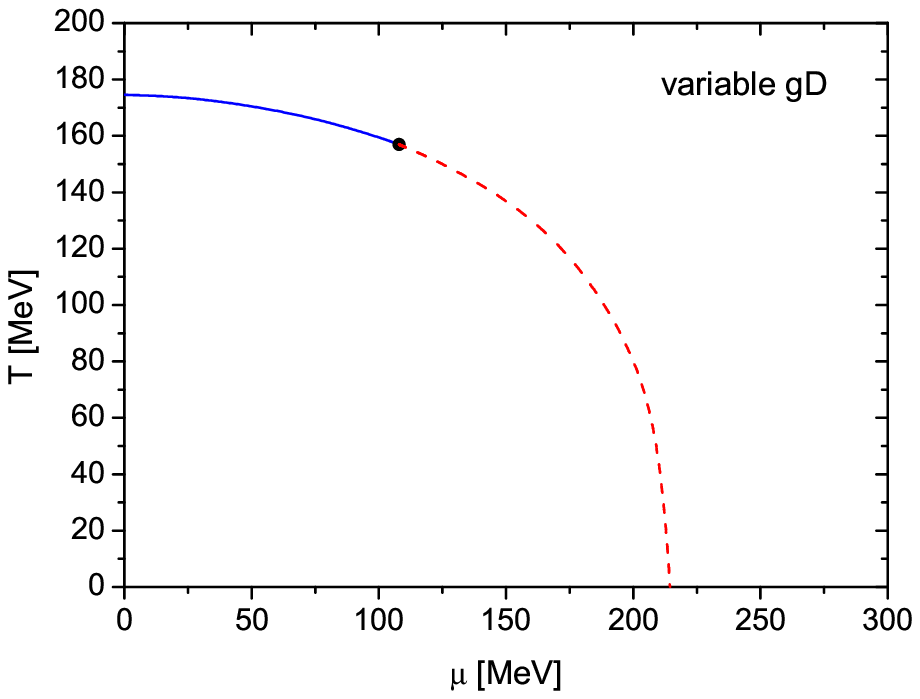,scale=0.7}
\epsfig{file=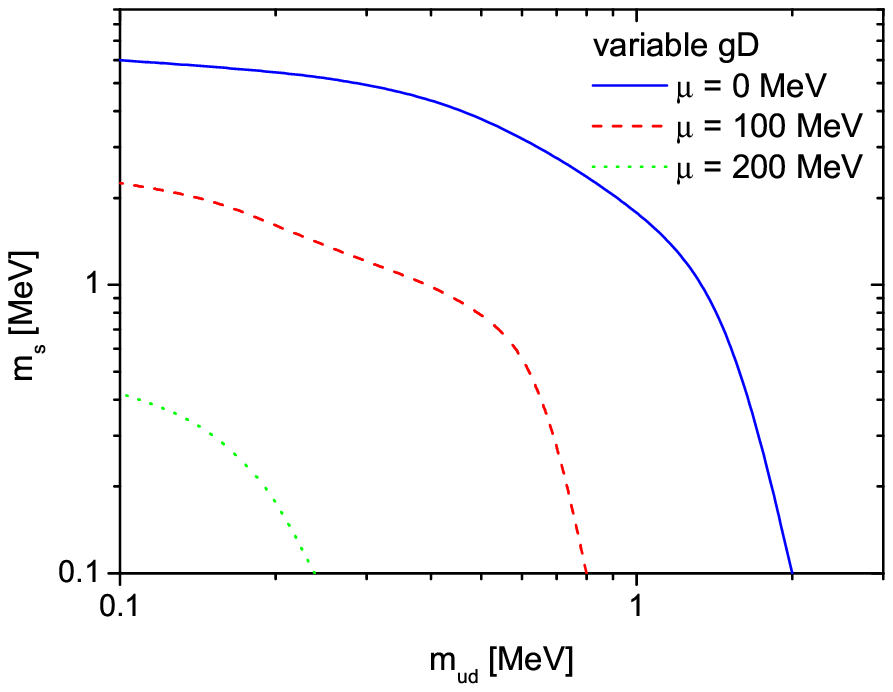,scale=0.7}
\end{minipage}
\begin{minipage}[t]{16.5 cm}
\caption{Left: Phase diagram of the 3-flavor PNJL model 
for unphysically small current quark masses $m_u=m_d=m_s = 0.5$~MeV
and a $\mu$-dependent six-point coupling constant
$g_D(\mu) = g_D^{(0)} \exp(-\mu^2/\mu_0^2)$, where $\mu_0 = 168$~MeV. 
Right: corresponding critical lines for $\mu = 0$, 100~MeV and 200~MeV 
in the plane of nonstrange and strange current quark masses. 
The first-order (crossover) regime is to the lower left (upper right)
of these lines.
\label{fig:gdvar}}
\end{minipage}
\end{center}
\end{figure}

It is now widely accepted that the finite-temperature QCD
``phase transition'' at $\mu = 0$ is in fact a rapid, but smooth
crossover. More precisely, this is expected to be the case in the
``real world'' with physical masses for up, down, and strange quarks.
For three massless flavors, on the other hand, the phase transition 
is known to be first order. Thus, if we take the nonstrange and 
strange quark masses as two independent variables, there is
a critical line in the $m_{ud} - m_s$ plane, which separates the
first-order regime at low masses from the crossover regime at higher
masses (see right panel of Fig.~\ref{fig:gdvar}).
On this line, the phase transition is second order.

When we now introduce the chemical potential as a third dimension,
the critical line becomes a critical surface. Until recently, it was 
generally expected that the critical surface is curved towards 
higher masses, so that the real world, which is located in the
crossover regime at $\mu=0$, may enter the first-order regime above
some critical value of $\mu$. A behavior like this is, for instance, 
seen in the NJL phase diagram of Fig.~\ref{fig:taylorpd} and in many
other model calculations. It is therefore often called the 
``standard scenario''. 
There are, however, indications from the lattice for a 
``non-standard scenario'', where the critical surface is curved
towards smaller masses \cite{deForcrand:2006pv}. 
If this was confirmed, it would mean that the real world always stays
in the crossover regime when $\mu$ is increased.
This by itself does not look unusual and can be realized easily in 
NJL and PNJL models (see below).
On the other hand, at very small quark masses, where the phase 
transition is first order at $\mu = 0$, the non-standard scenario would 
have the interesting consequence that there is a first-order phase 
transition at low values of $\mu$ which turns into a crossover
above some critical value $\mu$. 
Thus the question arises, whether such a ``reversed'' behavior is
something really exotic or whether it can be realized in a simple
model as well. 
In this context, it was recently suggested by Fukushima that the 
non-standard scenario could be realized within a PNJL model
if a $\mu$-dependent determinant interaction is introduced 
\cite{Fukushima:2008wg}.
Below, we work out this idea explicitly.

Following Ref.~\cite{Fukushima:2008wg} we consider a three-flavor
PNJL model, given by the Lagrangian
\be
   {\cal L} = \bar\psi(iD\hspace{-2.5mm}/ - \hat m)\psi
      + {\cal L}_4 + {\cal L}_6 - {\cal U}(\Phi,\bar\Phi)\,,
\ee
where $\psi$ is a quark field which is coupled through the covariant
derivative to the Polyakov loop. 
The terms ${\cal L}_4$ and ${\cal L}_6$ denote local 4-point and
6-point interactions, while ${\cal U}(\Phi,\bar\Phi)$ is the 
Polyakov loop potential. Most details of these terms are not important
for the following discussion and can be found in Ref.~\cite{Fukushima:2008wg}.
At this point we just mention that ${\cal L}_6$ is a determinant in 
flavor space (`` 't Hooft interaction''), which is proportional to
a coupling constant $g_D$.

\begin{figure}[tb]
\begin{center}
\begin{minipage}[t]{16.5 cm}
\epsfig{file=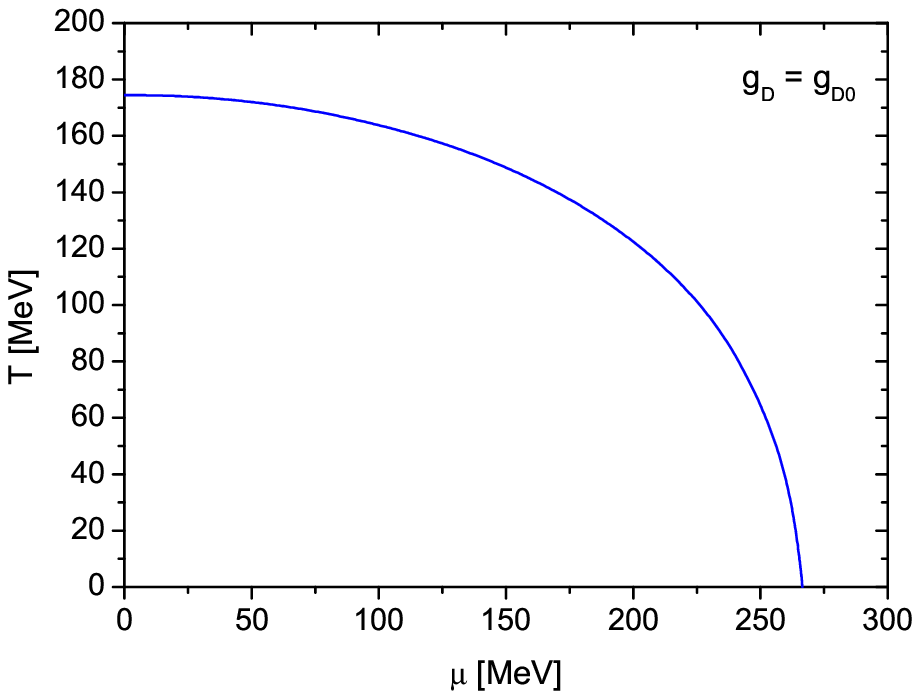,scale=0.55}
\hspace{-0.8cm}
\epsfig{file=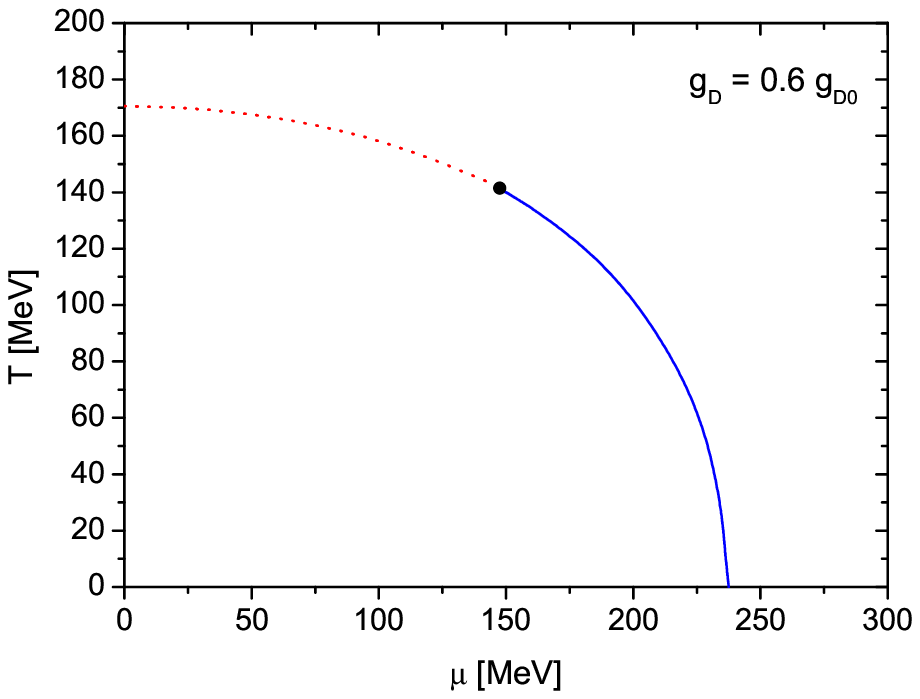,scale=0.55}
\hspace{-0.8cm}
\epsfig{file=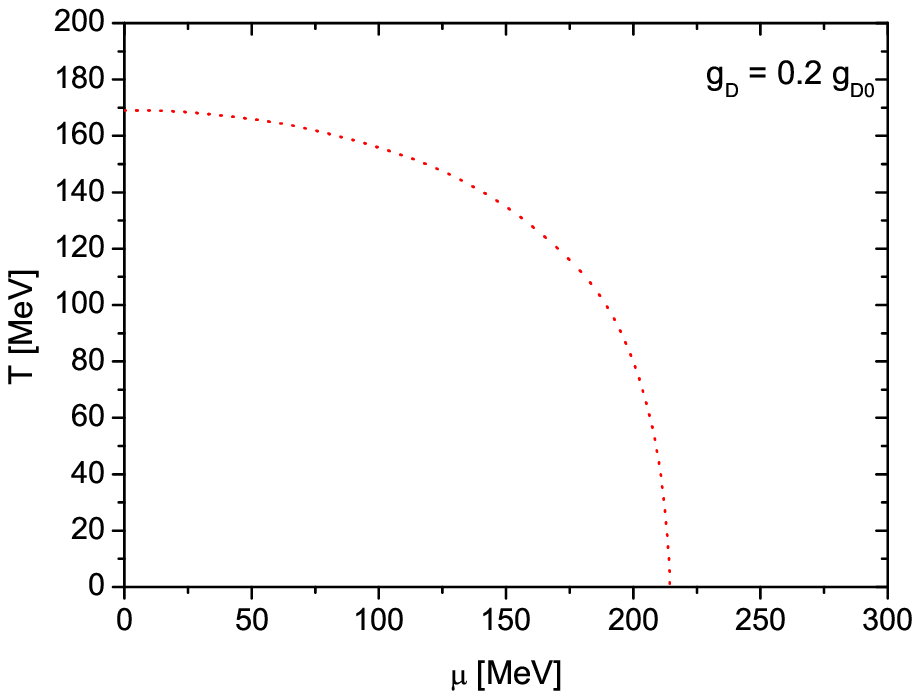,scale=0.55}
\end{minipage}
\begin{minipage}[t]{16.5 cm}
\caption{Phase diagrams of the 3-flavor PNJL model
for unphysically small current quark masses $m_u=m_d=m_s = 0.5$~MeV
and different values of the six-point coupling constant:
$g_D = g_D^{(0)}$ (left), $g_D = 0.6~g_D^{(0)}$ (center), and
$g_D = 0.2~g_D^{(0)}$ (right). The solid and dotted lines indicate
first-order phase transitions and crossovers, respectively.
\label{figphasegd}}
\end{minipage}
\end{center}
\end{figure}

Starting with a physical parameter set \cite{Fukushima:2008wg}
one finds a phase diagram which is qualitatively similar to 
Fig.~\ref{fig:taylorpd}, indicating that the model supports the standard 
scenario. In the next step we strongly decrease the
current quark masses to enter into the first-order regime at 
$\mu = 0$. The resulting phase diagram for 
$m_u = m_d = m_s = 0.5$~MeV is shown in the left panel of 
Fig.~\ref{figphasegd}. In this case, the phase transition is 
first order everywhere. Next, we decrease the value
of the determinant coupling from the original value $g_D^{(0)}$ to
smaller values. The resulting phase diagrams for $g_D = 0.6~g_D^{(0)}$
and $g_D = 0.2~g_D^{(0)}$ 
are shown
in the two other panels of Fig.~\ref{figphasegd}. Obviously, with
decreasing $g_D$ the first-order transition becomes gradually replaced
by a crossover and eventually disappears completely. However, in particular
the phase diagram in the center still looks ``normal'', with a first-order
transition at higher $\mu$ and a crossover at lower $\mu$. 
Nevertheless, the series of phase diagram displayed in Fig.~\ref{figphasegd}
shows that the reversed situation could be realized if we introduce a
coupling which drops sufficiently fast with $\mu$ \cite{Fukushima:2008wg}.
This is explicitly demonstrated in the left part of Fig.~\ref{fig:gdvar},
where we have chosen an exponentially dropping coupling constant,
$g_D(\mu) = g_D^{(0)} \exp(-\mu^2/\mu_0^2)$.
The constant $\mu_0 = 168$~MeV was simply introduced by hand to see the
desired effect and has no deeper meaning so far.

As argued above, the reversed form of the phase diagram shown in the
left part of Fig.~\ref{fig:gdvar} is a hint for a non-standard critical 
surface. This is confirmed by the results shown in the right part of 
Fig.~\ref{fig:gdvar}, where the critical lines in the 
$m_{ud} - m_s$ plane are shown for $\mu =$ 0, 100~MeV and 200~MeV.
We see that the critical surface is indeed curved towards lower masses
in this case. 
This example demonstrates that the non-standard scenario can be realized
in a relatively simple model. 
Moreover, since the determinant interaction is related to instanton effects, 
which are known to decrease with increasing density,  
the choice of the $\mu$-dependent $g_D$ appears quite natural.
Of course, it remains to be investigated whether our parameterization of
$g_D(\mu)$ makes sense also quantitatively. Furthermore, we should introduce
a $T$-dependence of $g_D$ as well. In fact, this could have the opposite 
effect.

\bigskip
{\large \bf Acknowledgments}
\\[1mm]
This talk was partially based on common work with D. Blaschke, M.K. Volkov, 
and J. Wambach, who we wish to thank for a fruitful collaboration.
Financial support by the Heisenberg-Landau programme (M.B. and A.E.R.),
by the grant of the Russian President (A.E.R.),
by CONICET and EMMI (A.G.G.) and by DFG (M.B.) are gratefully acknowledged.

\end{document}